# Schmidt decomposition of mixed-pure states for $(d,\infty)$ systems and some applications.

Roman Gielerak
University of Zielona Gora
Institute of Control&Computation Engineering
r.gielerak@ issi.uz.zgora.pl

***Summary***. *A simple derivation of finite Schmidt decomposition of pure states describing finite dimensional systems interacting with the infinite dimensional ones is presented. In particular, maximally entangled pure states in such systems are being characterized. The concept of mixed-pure states has been introduced and some criterions for checking separability and entanglement of them are presented .The notion of LOCC equivalence and LOCC semi-order on the space of pure states of systems analyzed has been adopted suitably. In particular a Nielsen -like theorem has been extended to the pure $(d, \infty)$- states case. The notion of spin-orbit entanglement in the context of atomic physics is being discussed from a mathematical perspective.*

## 1. Introduction

Quantum correlations contained in quantum entangled states describing composite quantum systems are by no doubts one of the major resources for several quantum information tasks [1,2,3,4,5].One of the very interesting feature of the quantum entanglement is its monogamy nature which means , roughly, that a total available amount of quantum correlations contained in quantum states is always of a limited capacity, depending on the very nature of system considered. In particular ,there exist quantum states containing maximal possible amount of quantum correlations and exactly these states are often most wanted for performing several quantum protocols like : teleportation of states , cryptographic protocols implementations and many others. This is the main reason that the mathematical description and the corresponding engineering technologies of preparing physically such maximally entangled states seems to be of great importance.

In the case of composite systems being in the maximally entangled state it is impossible further to entangle such quantum system with another quantum system , this monogamy principle is the major element that defends the security of the most of the implemented technically cryptographic protocols up to date [ 6].

In the case of finite dimensional systems a lot of work has been done on the very nature of quantum entanglement [1,2,3,4] , the case of two-partite systems is the best recognised situation. In the case of two-partite , finite dimensional systems , the Schmidt decomposition of the corresponding pure state gives essentially all relevant information on the corresponding quantum correlations. From the quantitative point of view and from qualitative ( (S)LOCC type, semi-order relations [2,3,4] ) point of view as well. The case of many-partite systems ,and also the case of mixed ( even for two-partite systems) states is much less recognised despite to many efforts [2,3,4,5].

The case of two-partite systems composed from finite dimensional system coupled with some infinite dimensional one is being discussed in the present note. A simple construction of the finite Schmidt decomposition of the pure states and mixed-pure of such systems will be demonstrated here and some straightforward consequences of the derived decompositions will be formulated . In particular, a general form of maximally entangled pure states of such systems is being derived and the corresponding amount of entanglement contained and defined as the von Neumann entropy of the arising reduced density matrices is calculated .

The present day physics of spin-orbit entanglement effects is the main motivation for our material presented in this note. Spin–orbit (SO) coupling—the interaction between a quantum particle's spin and its momentum—is ubiquitous in physical systems. The role playing in the atomic and molecular systems by the spin –orbit coupling is very well known since the very beginning of quantum theory of matter[7,8 ]. In recent years some new effects of spin –orbit correlations have been observed , the very nature of which can be explained on the assumption on the presence of entanglement between spin degrees of freedom with the orbital part of the corresponding wave functions. For more information on this and references see [9,10,11,12]. Also in condensed matter systems, SOE coupling and entanglement is crucial for the spin-Hall effect and topological insulators, and is important for spinotronic devices which might to appear to be of major importance in the future quantum computers hardware industries. Quantum many-body systems of ultra-old atoms forming new state of matter , the so called Bose-Einstein condensates can be today precisely controlled experimentally, and would therefore to provide an ideal platform on which one can study SOE coupling effects [9,10,11,12].

The organization of the material is the following one. In the next section we present very simple construction ( and presumably well known to the experts in the field ) of the finite Schmidt decomposition of pure states in the ( d,∞ ) class of two-partite systems. In particular the maximally entangled pure states of such systems have been described. In section3 the notation of the mixed-pure states for systems under consideration will be formulated and the corresponding canonical Schmidt decomposition presented and some applications are discussed. In particular several well known criterions for detection of entanglement in a such mixed-pure states are being adapted , the realignment criterion for detecting entanglement is among them. Section 4 contains ideas for extending the notion of LOCC semi-order on the space of mixed - pure states . The Nielsen theorem for the pure states of ( d, ∞ ) systems has been derived as the main result there.

In section 6 some computational aspects of the derived Schmidt decompositions is being discussed. In particular , the case of ½-spin states of Bloch type and the relativistic invariant of Dirac are discussed from the computational points of view. Also the monogamy of quantum entanglement is illustrated ,particular example of a single qubit entangled with the rest of the world described by some separable Hilbert space is discussed in more details .A very interesting hypothesis about absence of any entanglement in the case of mixed-pure states for ( 2, ∞ ) systems is formulated there.

## 2. The Schmidt decomposition theorem for pure states of (d, ∞) systems.

Let $h_1$, $h_2$ be a pair of separable Hilbert spaces, then the tensor product space $H = h_1 \otimes h_2$ is again separable Hilbert space with the corresponding scalar product

$$< -|->_H = <-|->_1^* <-|->_2$$

for product vectors in H and then extended by linearity and continuity to the whole of H.

The set of pure states $\partial E$ (H) on the space H is defined as the stratification of the action of unitary group U (1) ( the global phase calibration ) on the unit sphere in H.The space of quantum states E (H) on H is defined as the intersection of the cone of nonnegative , trace class of operators with the hyperplane tr ($\rho$) =1.From now on , we shall to assume that the factor $h_1$ is finite dimensional space , i.e. without loose of generality we assume that $h_1$ is the d-dimensional space $C^d$ with the standard Hilbert space structure.

For a given , separable Hilbert space H , we denote by CONS (H) the space of complete orthonormal systems in H .

Let $H = h_1 \otimes h_2$ , dim ($h_1$) =d $< \infty$ and let { $e_\alpha$ : $\alpha$ =1 : d } $\epsilon$ CONS ($h_1$) , { $f_k$ : k=1,....} $\in$ CONS ( $h_2$ ) . Then the set { $e_\alpha \otimes f_k$ , $\alpha$ =1 : d ,k=1....} forms a complete orthonormal system in $C^d \otimes h_2$ .Therefore , taking any vector $\Psi$ $\epsilon$ H we have the following decomposition ( norm convergent ):

$$\Psi = \sum_{\alpha,k} \psi_{\alpha,k} e_\alpha \otimes f_k \qquad (2.1)$$

where $$\psi_{\alpha,k} = <\Psi \mid e_\alpha \otimes f_k >$$

and $$\sum_{\alpha,k} \mid \psi_{\alpha,k} \mid^2 = \parallel \Psi \parallel^2$$

Let us define the following linear map:

$$J(\Psi) : C^d \rightarrow h_2 \qquad (2.2)$$

$$\sum_{\alpha=1}^{d} c_\alpha * e_\alpha \rightarrow \sum_{\alpha=1}^{d} c_\alpha * J(\Psi)(e_\alpha) =$$

$$\sum_{\alpha=1}^{d} c_\alpha \sum_{j=1}^{\infty} \psi_{\alpha,j} \mid f_j >$$

$$\sum_{j=1}^{\infty} (\sum_{\alpha=1}^{d} c_\alpha * \psi_{\alpha,j}) \mid f_j >$$

With the use of the Cauchy- Schwartz inequality it is not difficult to prove that the map $J(\Psi)$ is a bounded linear map and

$$\| J(\Psi) \|_{L(C^d \to h_2)} = \sup_{\|x\| \leq 1} \| J(\Psi)(x) \| \leq \| \Psi \| \tag{2.3}$$

Therefore the conjugate map

$$\| J^+(\Psi) \| : h_2 \to C^d \text{ is defined well by the identity :}$$

$$< F | J(\Psi)x > = < J^+(\Psi)F | x >$$

for all $x \in C^d$ and $F \in h_2$.

Defining

$$\Delta(\Psi) = J^+ * J(\Psi) : C^d \to C^d \tag{2.4}$$

it follows easy

(i)     $\Delta(\Psi) = \Delta^+(\Psi) \geq 0$

(ii)     $\| \Delta(\Psi) \|_{HS} = tr_{C^d} \Delta(\Psi) = \| \Psi \|$.

From the spectral theorem for hermitean matrices it follows the following spectral decomposition of the matrix $\Delta(\Psi)$:

$$\Delta(\Psi) = \sum_{\alpha=1}^{d} \tau_\alpha * | f_\alpha > < f_{\alpha|} | \tag{2.5}$$

where $\tau_\alpha \geq 0$ are the the corresponding eigenvalues of $\Delta$ and $f_\alpha$ are the corresponding eigenvectors. It also assumed the the appriopriate orthonormalisation procedura is performed ( in the case ofdegenerated eigenvalues ) that leads to orthogonality of the system of eigenvectors $f_\alpha$ .

The operator $J(\Psi) : C^d \to h_2$ being bounded is automatically closed and therefore the well known polarisation decomposition theorem can be applied [13,14] as well. This leads to the existence of some isometric map $V : Ker(J(\Psi)) \to Rank(J(\Psi))$ and such that the following equality holds :

$$J(\Psi) = V * \sqrt{\Delta(\Psi)} . \tag{2.6}$$

Using the decomposition ( 2.4) and comparing with the definition (2.2) we are seeing that :

$$\Psi = \sum_{\alpha=1}^{d} \tau_\alpha * |f_\alpha><g_\alpha| \tag{2.7}$$

where $g_\alpha = V|f_\alpha>$.

Thus we have proved the following result.

***Theorem*2.1 ( Schmidt decomposition theorem ).**

Let $H=C^d \otimes h$, where $h$ is a separable Hilbert space. Then, for any $\Psi \in H$ there exists an unique sequence of numbers $\tau_\alpha$, $\alpha=1{:}d$ and two CONS $\{f_\alpha\}$ in $C^d$ and $\{g_k\}$ in $h$ such that :

$$\Psi = \sum_{\alpha=1}^{d} \tau_\alpha |f_\alpha>|g_\alpha> \tag{2.8}$$

The crucial point here is that the corresponding Schmidt decomposition is finite, in particular, for any $\Psi$ the Schmidt rank is always finite and its maximal possible value is equal to d. The coefficients $\tau_\alpha$ defined in the unique way for a given $\Psi$ are called Schmidt coefficients of $\Psi$.

A straightforward consequence of this decomposition Theorem is the following result (being well known in the case of ( d , d' )-systems [15] ).

**Corollary 2.2.**

Let $H=C^d \otimes h$, where $h$ is a separable Hilbert space and $d{<}\infty$. Then for any vector $\Psi \in H$ the reduced density matrix $\rho_{c^d}(\Psi)$ of $\Psi$ after tracing out the degrees of freedom contained in $h$ are given by the formula :

$$\rho_{C^d}(\Psi) = tr_h(|\Psi><\Psi|) = \begin{bmatrix} \tau_1^2 & \cdots & 0 \\ \vdots & \ddots & \vdots \\ 0 & \cdots & \tau_d^2 \end{bmatrix} \tag{2.9}$$

where $\tau_i$ are the corresponding Schmidt coefficients of the vector $\Psi$.

Among different notions [2,3,4,5] of quantitative measures of amount of entanglement contained in pure quantum states we propose here, for an illustrative purposes mainly, the standard entropic measure Ent based on the notion of von Neumann entropy of the density matrix:

$$Ent(\Psi) = -tr\rho_{c^d}(\Psi)\ln(\rho_{c^d}(\Psi)) \tag{2.10}$$

Using the result ( 2.9 ) we have :

$$Ent(\Psi) = -\sum_{\alpha=1}^{d} \tau_\alpha^2 * \ln \tau_\alpha^2. \tag{2.11}$$

As is well known the maximum value of this function is attained for the case $\tau_\alpha = 1/\sqrt{d}$, for all $\alpha$. In this way we have proved the following Corollary .

**Corollary 2.3**

*Let $H = C^d \otimes h$ , where $h$ is a separable Hilbert space and $d < \infty$ .Any maximally entangled pure state $\Psi$ of the system described by $H$ is necessary of the form*

$$\Psi = \sum_{\alpha=1}^{d} \frac{1}{\sqrt{d}} \psi_\alpha \otimes f_\alpha \qquad (2.12)$$

*where $\{\psi_\alpha , \alpha = 1{:}d \} \in CONS ( C^d )$ and $\{ f_\alpha , \alpha = 1{:}d \}$ is d-dimensional orthonormal system of vectors in h.*

These results can also be extended to the case of the spaces $L_2(C^d) \otimes h$ ,
and also (up to some extent of course ) to the more interesting situation of $L_2(C^d) \otimes L_2(h)$ space , here $L_2(h)$ refers to the Hilbert -Schmidt structure on the space $L(h)$ of bounded, linear operators on h.

## 3. **The case of mixed-pure states.**

The natural Hilbert space structure in the space of endomorphisms of the space $C^d$ , for $d < \infty$ is that given by the Hilbert-Schmidt scalar product : for $A, B \in L(C^d)$ we define

$$< A \mid B >_{HS} = tr A^+ B \qquad (3.1)$$

The space of mixed states of a d –dimensional quantum system, denoted as $E(C^d)$ is given by

$$E(C^d) = \{\rho : \rho^t = \rho \geq 0, \ tr\rho = 1\} \qquad (3.2)$$

The space of mixed-pure states of a bipartite system described by the tensor product $C^d \otimes h$ , where h is a separable Hilbert space ,
is defined as

$$M - pE(C^d \otimes h) = \{ Q \ \varepsilon \ HS (C^d) \otimes h : Q^t = Q \geq 0 , tr (Q) = 1,$$
$$tr_h (tr_{C_d} (Q) )^2) = 1 \} \qquad (3.3)$$

In particular , the set of M-p states of the form $Q = q \otimes \Psi , q \ \varepsilon \ E(C^d)$ , $\Psi \ \varepsilon$ h is normalized vector in h,, is called the space of mixed-pure factor states and will be denoted as Mp-F( $C^d \otimes h$ ) . The , finite convex envelope of the set Mp-F will be called the set of Mp-separable states and denoted as Mp-Sep ( $C^d \otimes h$ )

$$Mp - Sep(C^d \otimes h) = \{Q \varepsilon HS(C^d) \otimes h : \exists$$
$$finite \ decomposition \ Q = \sum_i \rho_i \otimes \psi_i ,$$
$$\rho_i \in HS(C^d), \psi_i \in h\} \ \cap M - pE \qquad (3.4)$$

**Definition 3.1**

A state $Q \in$ M- pE( $C^d \otimes$h ) is called Mp-entangled state iff

$$Q \in M\text{-pE} ( C^d \otimes h ) \backslash \text{closure of Mp-Sep} (C^d \otimes h )$$

As in the standard , finite dimensional ( d, d' ),d`< ∞ situation ,no necessary and sufficient conditions for the presence of Mp-entanglement are known for the mixed-pure states . However , several sufficient conditions for detecting entanglement are naturally adopted to the case at hand . The one , known in the standard situation as Entanglement Realignment Criterion ( see i.e. [23,24] ) especially straightforwardly is adapted to the case of M-p states. For this , let as note the following corollary of Theorem 2.1.

**Proposition 3.1**
*Let h be a separable Hilbert space and let $d < \infty$ .Then , for any $Q \in HS ( C^d ) \otimes h$ there exists a sequence of nonnegative numbers $\lambda_{\alpha\beta}$ , a family $(E_{\alpha\beta}) \varepsilon CONS ( HS ( C^d ) )$ and a orthonormal family $(\psi_{\alpha\beta} , \alpha,\beta =1:d )$ in h such that the following Schmidt decomposition holds :*

$$Q = \sum_{\alpha,\beta=1}^{d} \lambda_{\alpha\beta} E_{\alpha\beta} \otimes \psi_{\alpha\beta} \qquad (3.5)$$

As it follows from the results of [25,26,28] for the case of hermitean Q the orthonormal system $E_{\alpha\beta}$ can be seen as consisting of hermitean matrices. If additionally these matrices are nonnegative for the case of nonnegative Q it follows, that performing canonical Schmidt decomposition we have proved that the object Q is separable in the sense (and therefore hermitean) of definition (3.4).

**Corollary 3.2**
*Let $Q \varepsilon Mp\text{-}E( C^d \otimes h )$ . If the corresponding matrices $E_{\alpha\beta}$ in the decomposition ( 3.5) for Q are nonnegative then Q is Mp-Sep ( $C^d \otimes h$ ) state.*

Now we are ready to formulate one of the main result of this section.

**Theorem 3.2**
*Let $Q \in M\text{-}pE(C^d \otimes h )$ and let $(\lambda_{\alpha\beta} , \alpha,\beta =1:d )$ be the corresponding Schmidt coefficients of Q as given in Proposition 3.1.*
*If*

$$\sum_{\alpha,\beta=1}^{d} \lambda_{\alpha\beta} > 1$$

*then the state is Mp-entangled .*
Proof:
A well known technique of entanglement witness could be adapted to the situation considered here. For this, let

$$Q = \sum_{\alpha,\beta=1}^{d} \lambda_{\alpha\beta} \, E_{\alpha\beta} \otimes \psi_{\alpha\beta}$$

be the Schmidt decomposition of the mixed - pure state Q as given by proposition 3.1. Let us define the following operator ( operator-vector, depending on the particular situation )

$$W_Q = \sum_{\alpha,\beta=1}^{d} E_{\alpha\beta} \otimes \psi_{\alpha\beta} \quad (3.6)$$

*Observation (1)*:
For all Q' ∈ M-pSep ( C$^d$⊗h) ) the following is valid

Tr Q' *W$_Q$≥ 0

Proof: It is enough to assume the Q' is of the form Q' = q ⊗ψ,q∈ E(C$^d$), ψ∈h, ∥ ψ ∥ = 1. Then, using ( 3.6):

$$Tr(Q'W_Q) = 1 - \sum_{\alpha,\beta=1}^{d} tr_{C^d}(E_{\alpha\beta} \cdot q) * tr_h(\hat{\psi}_{\alpha\beta} \cdot \hat{\psi})$$

where
$$\hat{\psi} = | \psi >< \psi |, \hat{\psi}_{\alpha\beta} = | \hat{\psi}_{\alpha\beta} >< \hat{\psi} |.$$

From the Cauchy-Schwartz's inequality it follows that:

$$\sum_{\alpha,\beta=1}^{d} tr_{Cd}(E_{\alpha\beta} \cdot q) tr_h(\hat{\psi}_{\alpha\beta} \cdot \hat{\psi})$$
$$\leq (\sum_{\alpha,\beta=1}^{d} (tr_{Cd}(E_{\alpha\beta} \cdot q))^2)^{1/2} * (\sum_{\alpha,\beta=1}^{d} (tr_h(\hat{\psi}_{\alpha\beta} \cdot \hat{\psi}))^2)^{1/2} \leq 1 \quad (3.7)$$

Because ( E$_{\alpha\beta}$ )∈CONS( HS( C$^d$ ) ) and ∥ρ∥$_{HS}$ = 1 and additionally |tr ($\hat{\psi}_{\alpha\beta} \cdot \hat{\psi}$ )| ≤ 1 for all α, β.

*Observation (2 )* :
The following inequality is true: for Q given by (3.5)

Tr ( Q W$_Q$ ) < 0
Proof :

$$Tr_{WQ} \cdot Q = 1 - \sum_{\alpha,\beta=1}^{d} \lambda_{\alpha\beta} tr_{C^d} E_{\alpha\beta} \cdot E_{\alpha'\beta'} * tr_h(\hat{\psi}_{\alpha'\beta'} \hat{\psi}_{\alpha\beta}) = 1 - \sum_{\alpha,\beta=1}^{d} \lambda_{\alpha\beta} \quad (3.8)$$

from the orthonormality of the system $(\psi_{\alpha\beta})$. So, if $\sum_{\alpha,\beta=1}^{d} \lambda_{\alpha\beta} > 1$ the result follows.

Collecting these two observations together we conclude that the proof of Theorem is completed.
**q.e.d.**

Also M-p separablity questions can be asked with the help of the Schmidt decomposition ( 3. 5 ). According to the methods presented in [25,26,27,28] the following theorem can be proved ( the details, being in fact straightforward ( and this is connected to the question how far we are from nonnegative matrices E $_{\alpha\beta}$ in the Schmidt expansion (3.5) ) extensions of the corresponding methods to the case at hand will be presented elsewhere[16].

Let us define the Schmidt rank of the state Q M-pE ( $C^d \otimes$ h) as a number of non-vanishing Schmidt coefficients of the corresponding decomposition (3.5 )

**Theorem 3.3**
    If the Schmidt rank of a state Q $\in$ M-pE ($C^d \otimes$ h) is less or equal to 2 then the state Q is separable .

**Remark.**
    Many of other, entanglement detecting criterions ,well known in the context of ( d, d' ) d*d'<$\infty$ systems can be easily adapted to the considered here Mp-states of (d, $\infty$) systems.
    As an example let us consider the partial transpose positivity condition (PPT condition) introduced by Peres and then developed in many papers [27,28,1,2,3]. For this goal let us define the following partial transposition operation

$$T_L = HS(C^d) \otimes h \to HS(C^d) \otimes h$$
$$\rho \otimes \psi \to \rho^t \otimes \psi$$
$$(3.9)$$

and extended by linearity and continuity to whole space HS($C^d$)$\otimes$h.

For Q $\in$ Mp-Sep ($C^d \otimes$h) we have:

$$T_L(Q) \geq 0.$$

So, if $T_L(Q) < 0$

$$(3.10)$$

then $Q \in Mp\text{-}Ent \ (C^d \otimes h).$

    The minimal dimension of the nontrivial space HS($C^d$)$\otimes$h corresponding to the values d=dim h=2 is equal to 8. In terms of the canonical Schmidt decomposition this criterion means :

if, $Q = \sum_{\alpha,\beta=1}^{d} \lambda_{\alpha\beta} E_{\alpha\beta} \otimes \psi_{\alpha\beta} \geq 0$          (3.11)

then $T_L(Q) = \sum_{\alpha,\beta=1}^{d} \lambda_{\alpha\beta} E_{\alpha\beta}^{t} \otimes \psi_{\alpha\beta}$.          (3.12)

So, if $E_{\alpha\beta}^{t} = E_{\alpha\beta} \geq 0$ then $T_L(Q) \geq 0$, and if Q is expanded in terms of nonpositive hermitian matrices the PPT-criterion does not works in general case.

# 4. LOCC and partial ordering of pure states.

Bipartite, pure state entanglement manipulations are relatively well understood presently in the case of bi-partite, finite- dimensional systems. In particular, the class of LOCC operations is coincident with the class SEP in this case [17]. Basing on the above mentioned fact, together with the theory of bistochastic matrices a complete description of the LOCC semi-order on the space of pure states can be translated into the corresponding semi-order relations on the space of the Schmidt coefficients as was shown by Nielsen [18] and developed further in many papers[2,3,4]. The very interesting problem here is to extend these results to the case of ( d, $\infty$ ) systems considered here, see [16].\

Let $L_{fd}(h)$ be a lattice of finite dimensional subspaces of h. For a given set $\Delta = \{\psi_1, ..., \psi_k\}$ of vectors in h the linear space generated by them will be denoted as $L(\Delta) = lh\{\psi_1, ..., \psi_k\}$. For a given subspace $\Sigma \, \varepsilon \, L_{fd}$ let us denote the algebra of linear operators acting essentially in $\Sigma$ :

     $LB(\Sigma) = \{ A : \Sigma \rightarrow \Sigma : $ linear and $A_{|\Sigma^c} = 1, A \, \varepsilon \, End(\Sigma)\}$

**Lemma 1 .**
*Let $\Sigma \, \varepsilon \, L_{fd}$, $dim(\Sigma) = d$. Then $LB(\Sigma)$ is unitary isomorphic with $End(C^d)$.*
Proof:
All d-dimensional Hilbert spaces are unitary isomorphic as we know from the basic unitary spaces theory. In particular, it follows from this remark that $LB(\Sigma)$ could be unitarly identified with $C^d$. **q.e.d**.

Let $\Sigma \, \varepsilon \, L_{fd}$, $dim(\Sigma) = d$. The space of states $E(\Sigma) = \{ \rho \, \varepsilon \, LB(\Sigma) : \rho \geq 0$ and $tr_{\Sigma}(\rho) = 1 \}$ is then isomorphic to the space $E(C^d)$ as it follows by similar arguments. In particular the set of pure states on $\Sigma$, $P(\Sigma) = \{\Sigma \, \varepsilon \, E(\Sigma) : tr(\Sigma^2) = 1 \}$ is identical with that on the space $C^d$ and consists of orthogonal projectors onto the unit vectors of $\Sigma$ ( modulo the action of global $U(1)$ group ).

By a $\Sigma$- based projective measurement we will mean a finite sequence of operators $F_i \, \varepsilon \, LB(\Sigma)$, i=1:n and a sequence of reals $p_i > 0$ such that :
( i ) $\sum_{i=1}^{n} F_i^{+} F_i = 1$
(ii ) $\sum_{i=1}^{n} p_i = 1$
The measurement consists of transforming the actual state $\rho$ of the system onto the state $F_i^{+} \rho \, F_i$ with probability $p_i$ .

The set of projective measurements on $\Sigma$ will be denoted $pM(\Sigma)$ and this set is obviously unitary isomorphic set of projective measurements $pM(C^d)$.

**Lemma 2 .**
*Let $\Sigma \varepsilon L_{fd}$, dim $(\Sigma) = d$. Then the set $pM(\Sigma)$ is unitary isomorphic with the set of projective measurements on $E(C^d)$.*

Let $S_d = \{ (\lambda_i , \text{ for } i=0{:}d ) : \lambda_i \varepsilon [0,1] \}$ be d-dimensional simplex in $R^d$. Then , after Nielsen [2,18] , and followers [3,4] , we define certainpartial order in $S_d$ by the following definition

(1) for a given $\lambda$ let us perform the ordering of its component in decreasing order obtaining $\lambda \varepsilon S_d$ such that ,

for all $i=0{:}d-1 \lambda_i >= \lambda_{i+1}$ .

(2) For two points $\lambda^1$ and $\lambda^2$ of the simplex $S_d$ ordered as in point (1) above we say that

$\lambda^1 \preccurlyeq \lambda^2$ Wpisz tutaj równanie.
iff for all $i=0:d$

$$\sum_{j=1}^{i} \lambda_j^{1} \leq \sum_{j=1}^{i} \lambda_j^{2}$$

Now , let $H = C^d \otimes h$ , where $h$ is a separable Hilbert space and let $\Psi^1$, $\Psi^2 \varepsilon H$ . We will say that that the state $\Psi^2$ is LOCC majorised by the state $\Psi^1$ iff there exists a sequence of projective measurements $M^1 = (F^1_\alpha, p^1_\alpha)$ , and a sequence of local unitary trnsormations $M^2 = (1 \otimes F^2_\alpha)$ ,both supported on the finite dimensional subspaces of $h$ such that the state $\Psi^1$ can be transformed by the standard LOCC protocol , with the use of measurement $M^1$ and local unitary transformations $M^2$ onto the state $\Psi^2$ ,.together with local communication.

Theorem 4.1 ( extension of the Nielsen theorem )
*Let $h$ be a separable |Hilbert space and let $d$ be finite. Then , the pure state $\Psi^1 \varepsilon P (C^d \otimes h)$ can be transformed by the local , finite dimensional operations as above to the state $\Psi^2 \varepsilon P (C^d \otimes h)$ iff*
$\lambda(\Psi^1) \preccurlyeq \lambda(\Psi^2)$

*where $\lambda(\Psi)$ stands for the corresponding Schmidt coefficients of the Schmidt decompositions of the pure state $\Psi$ .*

Proof:
    Let
$\Psi^\alpha = \sum_{i=1}^{d} \lambda^\alpha_i \ e^\alpha_i \otimes \psi^\alpha_i$ , for $\alpha = 1,2$ be the corresponding Schmidt decompositions of a given vectors $\Psi^\alpha \varepsilon P (C^d \otimes h)$ .

Let $J^\alpha$ be the isomorphism maps from the spaces $h^\alpha = lh \{ \psi^\alpha_i, i=1:d \}$ to $C^d$.

$$J^\alpha: \qquad \psi^\alpha_i \to f_i^{\;\alpha}$$

and extended by linearity to the whole space $h^\alpha$, where $\{ f^\alpha_i, i=1:d \}$ is any CONS in $C^d$.

Then any linear transformation A acting on $C^d$ can be transported to the space $h^\alpha$ by the formula $A^\alpha = (J^\alpha)^{-1} A J^\alpha$.

. In particular any projective measurement defined on the space $C^d$ can be transported to act as a projective finite dimensional measurement in h, in fact in the corresponding spaces $h^\alpha$. From the assumptions made on $\lambda$-as and from the original Nielsen theorem it follows that the vector $\Psi^{1\sim} = (id \otimes J^1)(\Psi^1)$ can be transformed onto the vector $\Psi^{2\sim} = (id \otimes J^2)(\Psi^2)$ with the use of some LOCC protocol together with the corresponding pair of consequent projective measurements $M^1$ and local unitary operations $M^2$. So, it is enough to transport by the similarity transformations $J^\alpha$ the actions of the corresponding measurements in $C^d$ onto the corresponding finite dimensional measurements on the spaces $h^\alpha$. **q.e.d.**

Remarks5.1 . Several results concerning (S)LOCC transformations and LOCC and LU unitary equivalences of pure states of ( d, $\infty$ ) system considered here , well known for the finite dimensional case [2,3,4] , can be extended to the case analysed here by asimilar adaptations of results established there.

# 5. Practical computations and some examples.

The presented here derivation of the Schmidt decomposition is hardly to be used for the practical , even computer assisted calculations due to the a priori infinite series to be summed up ( see (2.1).However, in some very special situations the needed numerical calculations can be performed by standard , widely available tools like SVD decomposition. The one of such situation do appear if the vector $\Psi$ is given by the finite sum of the form :

$$\Psi = \sum_{\alpha,k} c_{\alpha,k} e_\alpha \otimes f_k \; (5.1a)$$

where $e_\alpha \in C^d$ and $f_k \in$ h . Let r= rank $\{ e_\alpha \} \leq d$ and s= rank $\{ f_k \} < \infty$ be the ranks (i.e. dimensions of the linear subspaces in the corresponding vector spaces spanned by the corresponding sets of vectors ) of the arising system of vectors in the representation (5.1a). Then we can perform the Gram-Schmidt orthonormalisation operation obtaining a new systems $\{ e^*_\beta, \beta=1:r \}$ and $\{ f^*_k, k=1:s \}$ , this time we obtain the systems that are composed from orthogonal and normalised vectors and the vector $\Psi$ can be represented as

$$\Psi = \sum_{\alpha,k} D_{\alpha,k} e^*_\beta \otimes f_k^{\;*} \; (5.1b)$$

Completing (if necessary ) the orthonormal systems used in ( 2.14 ) we can apply   by suitable reorganisation the corresponding matrix D  the standard SVD decomposition method to compute the  Schmidt coefficients   of the vector $\Psi$.

The following  Theorem  may be  used  also  to  compute  some  approximations to the exact  values  of  rank  and  Schmidt  coefficients in the  considered here  case.

**Theorem  5.1**

*Let $\Psi_n$ be a norm convergent sequence of   vectors  in $H = C^d \otimes h$ , where  h is separable . Let SD = { $r_n$, {$\tau^n_\alpha$} } be  the  corresponding Schmidt data  for $\Psi_n$. Then*

$$\lim_{n \to \infty} r_n = r$$

*and*                         $\lim_{n \to \infty} \tau^n_\alpha = \tau_\alpha$ *for  $\alpha = 1 : d$*

*where( r, $\tau_\alpha$) are  the  Schmidt data  of  $\Psi$ .*

Proof :

From  the  linearity  of  the  map  J  and  the  estimate  ( 2.3)  it  follows  that  if $\Psi_n \to \Psi$ *in the norm* as  n $\to \infty$  *then* J ( $\Psi_n$) $\to J(\Psi)$  in the  norm. Therefore  also  J ( $\Psi$ )   tends  to  J ( $\Psi$ )  as n $\to \infty$  in the  norm  topology.  Thus, we have  proved  that

$$\lim_{n \to \infty} \Delta(\Psi_n) = \Delta(\Psi) \text{ in the norm of } L(C^d).$$

Let  us  order  the  eigenvalues  {$\tau^n_\alpha$, $\alpha$=1:d}  of  the matrices $\Delta(\Psi)_n$ and of  the  matrix $\Delta(\Psi)$ in the  decreasing  order.  Then, by  using  the  Weyl  theorem  [19]  we  have

$$\max_{1 < \alpha < d} | \tau^n_\alpha - \tau_\alpha | << \| \Delta(\Psi_n) - \Delta(\Psi) \| . \quad (5.2)$$

To  apply  this  theorem  to  the  computability  of  Schmidts  data  for  a  given  vector $\Psi$ given  by  the  infinite  dimensional  representation  like  (2.1 )  we  observe  that  any  such vector  can  rewritten  in  the  following  form:

$$\Psi = \sum_{\alpha,k} \psi_{\alpha,k} e_\alpha \otimes f_k = \sum_k (\sum_\alpha \psi_{\alpha,k} e_\alpha) \otimes f_k \qquad (5.3)$$

Let $P_N$  be  an  orthogonal  projection  on  the  subspace  $h_N$ = l.h { $f_1$ ,...,$f_N$ }  of  h . We  define $\Psi_N = P_N \Psi$, then $\Psi_N$ is  finite  dimensional  represented  and  $\Psi_N \to \Psi$ as  N$\to \infty$ in  the  norm . Therefore ,  we  can  apply  the  standard  SVD  decomposition  method computing  the  corresponding  Schmidt  data  for  the  sequence  of  vectors $\Psi_N$. Then  we analyse  behaviour  of  the  Schmidt  coefficients  of  vectors $\Psi_N$ having  hope  to  guess  the corresponding  limiting  values. As  the  corresponding  sequences  of  Schmidt  numbers  are convergent  as  it  follows  from  Theorem 3. they  are  automatically  Cauchy  type  sequences. In  any  case  by  this  method  we  can  compute  the  exact  value  of  Schmidt  rank  of $\Psi$ and  the  Schmidt  coefficients of $\Psi$ with  an  arbitrary  precision  in  a  finite  time.

## Example 5.1. The spin ½ Bloch wave functions.

Let $d=2$ .Then for any orthonormal system { $e_1$ ,$e_2$ } of vectors in $C^2$ and any CONS { $f_k$ , k=1,2,…} in a separable Hilbert space h=$L_2$ ($R^3$ ) and any normalised vector $\Psi$ ε $C^2 \otimes$ h we can write:

$$\Psi = \sum_{\alpha,k} \psi_{\alpha,k} e_\alpha \otimes f_k = c_1 * e_1 \otimes \theta_1 + c_2 * e_2 \otimes \theta_2 \qquad (5.4)$$

where

$$\theta_\alpha = \sum_{k=1}^{\infty} \psi_{\alpha,k} * f_k * \| \sum_{k=1}^{\infty} \psi_{\alpha,k} * f_k \|^{-1} . (5.5)$$

For α= 1,2 , $c_\alpha = (\sum_k |\psi_{\alpha,k}|^2)^{\frac{1}{2}}$ .

The index of similarity δ is defined as :

$$\delta = |1 - <\theta_1 | \theta_2 >|. \qquad (5.6)$$

**The case δ=1**.

In this case $<\theta_1| \theta_2> =0$ and then the expression ( 3.2) is exactly the Schmidt decomposition of $\Psi$ and the corresponding entropy of entanglement contained in $\Psi$ is given by :

$$Ent(\Psi) = -c_1^2 * \ln(c_1^2) - c_2^2 * \ln(c_2^2) \quad (5.7)$$

which attains its maximum for $c_\alpha = 1/\sqrt{2}$ and then

$$Ent(\Psi) = \ln(2).$$

**The case δ=0** .

In this case the vectors $\theta_1$ and $\theta_2$ are linearly dependent and the vector $\Psi$ is separable.

**The case 0< δ<1** .

In this case the vectors $\theta_1$ and $\theta_2$ are linearly independent and by applying the well known Gram-Schmidt procedure it is possible to rotate the vectors $\theta_1$ and $\theta_2$ into the system of orthonormal vectors { $\theta_1^*$ , $\theta_2^*$ } and such that :

$$\theta_1 = g_{11} \theta_1^* + g_{12} \ \theta_2^*$$

$$\theta_2 = g_{21} \theta_1^* + g_{22} \ \theta_2^* \qquad (5.8)$$

with $|g_{\alpha 1}|^2 + |g_{\alpha 2}|^2 = 1$ for α=1,2
and : $g_{11} = 1$, $g_{12} = 0$ ,$g_{21} = -\alpha/\beta$ , $g_{22} = 1/\beta$ , where

$$\alpha = \sqrt{\frac{1}{1-\sigma^2}}, \beta = \sqrt{\frac{1}{1-\sigma^2}}, \sigma = <\theta_1 \mid \theta_2 > (5.9)$$

Therefore, we can rewrite $\Psi$ as

$$\Psi = c_1 * e_1 \otimes \theta_1^* + c_2 * \sigma * e_2 \otimes \theta_2^* + \frac{c_2}{\beta} * e_2 \otimes \theta_2^* (5.10)$$

It is not difficult to compute the corresponding density matrix the qubit after tracing out the degrees of freedom corresponding to the environment in which the qubit is located:

$$\rho_{c^2} = tr_h < \Psi \mid \Psi >= \begin{bmatrix} c_1^2 & c_1 c_2 \sigma \\ c_1 c_2 \sigma^* & c_2^2 \end{bmatrix} \tag{5.11}$$

By an elementary ( but little bit tedious ) calculations of the corresponding eigenvalues of the matrix $\rho_{c^2}$ it follows that for $1 < \mid \sigma \mid \leq 1$

$$Ent(\rho_{c^2}) = -tr\rho_{c^2}\ln(\rho_{c^2}) < \ln(2). (5.12)$$

Therefore if $0 \leq \delta < 1$, then always Ent $(\Psi) < \ln(2)$.
Whether there exists a Bloch wave function corresponding to some realistic physical situation with maximal entanglement amount in between spin and orbital degrees of freedom is not known to the Author.

## Example 5. 2  The Dirac wave functions.

This corresponds to the fully relativistic situation d=4 and h = $L_2$ ( $R^4$ ) , i.e. H = $C^4 \otimes L_2$ ( $R^4$ ) .In this space unitary representation of the ( covering of ) Poincare group is realised and the corresponding wave functions are being expressed as :

$$\Psi = \begin{pmatrix} \psi_0 \\ \psi_1 \\ \psi_2 \\ \psi_3 \end{pmatrix} = \sum_0^3 e_\alpha \psi_\alpha \qquad (5.13)$$

and where { $e_\mu$ , $\mu$=0:3 } is the canonical basis of $C^4$ , ( $e_\mu$ )$_\alpha$ =$\delta_{\alpha\mu}$for $\alpha,\mu = 0:3$ and $\psi_\mu$ $\varepsilon$ $L_2$ ($R^4$) and such that
$\sum_{\mu=0}^3 \mid \psi_\mu \mid_{L_2}^2 = 1$.

Such type of wave functions are arise often in physics of relativistic electrons, in particular case as a solutions of the famous Dirac wave equations :

$$D\Psi = j \qquad\qquad (5.14)$$

where D is the Dirac operator given as $D = \gamma_\mu \delta_\mu + im$ , $\gamma_\mu$ are so called gamma matrices and j is some 4-current..

It seems of some interest to analyze the spin -orbit entanglement present in a particular Dirac spinor depending on the orthogonality relations in between the components of it.For this purpose let us define the Gram matrix $G(\Psi)$ for the spinor $\Psi$ in the following standard way $G(\Psi)_{\alpha\beta} = < \psi_\alpha$ I $\psi_\beta >_{L2}$.Let us start with the easiest case :

**Case 0**: Rank $(G(\Psi)) = 4$ , $G(\Psi)_{\alpha\beta} = \delta_{\alpha\beta} * c_\alpha$ , $c_\alpha > 0$, $\Sigma c_\alpha^2 = 1$

In this case the representation ( 5. 12 ) is exactly the Schmidt decomposition of the spinor $\Psi$ and then the corresponding entropy of spin-orbit entanglement contained in the spinor $\Psi$ is equal to

$$En(\Psi) = -\Sigma c_\alpha^2 * \ln (c_\alpha^2)$$

The maximum value of which is attained for the values $c_\alpha = 1/2$.
Our conjecture here is :

**Conjecture 5.1**
  The maximal value of entropy of spin-orbit entanglement in the case of Dirac spinors is equal to $\ln(4)$.The same is valid also for the case of possible mixed-pure spinor states.

Numerical experiments performed by us strongly supports the validity of this Conjecture [16].
Whether there exist a realistic situations which lead to the maximally entangled solutions of the Dirac equations (5.14) is not known to the Author.

# Example 5.3 Entanglement in a photonic states.

Similar analysis can be performed with the photonic wave functions , thus leading to the proper , from the mathematical point of view notion of possible entanglement of polarisation degrees of freedom and the orbital one [16] .This should be contrasted with somewhat fuzzy and unclear notion on such physical effects that is present in the current literature .

**Address :**

University of Zielona  Gora, Institute of  Control& Computing  Engineering, ul.Szafrana 2,65-516 ZielonaGóra , Poland

r.gielerak@issi.uz.zgora.pl